# Surface Sensitive NMR Detection of the SEI Layer on Reduced Graphene Oxide

*Michal Leskes[(1)]\*, Gunwoo Kim[(2,3)], Tao Liu[(2)], Alison L. Michan,[(2)] Fabien Aussenac[(4)], Patrick Dorffer[(4)], Subhradip Paul[(5)], Clare P. Grey[(2)]\**

[(1)]Department of Materials and Interfaces, Weizmann Institute of Science, Rehovot, 76100 Israel

[(2)]Department of Chemistry, University of Cambridge, Lensfield Road, Cambridge, CB2 1EW, UK

[(3)]Cambridge Graphene Centre, University of Cambridge, Cambridge, CB3 0FA, UK

[(4)]Bruker BioSpin, 34 rue de l'Industrie BP 10002, 67166 Wissembourg Cedex, France

[(5)]DNP MAS NMR Facility, Sir Peter Mansfield Magnetic Resonance Centre, University of Nottingham, Nottingham, NG7 2RD, UK

**\*michal.leskes@weizmann.ac.il, cpg27@cam.ac.uk**

Abstract

The solid electrolyte interphase (SEI) is detrimental for rechargeable batteries' performance and lifetime. Understanding its formation requires analytical techniques that provide molecular level insight. Here dynamic nuclear polarization (DNP) is utilized for the first time for enhancing the sensitivity of solid state NMR (ssNMR) spectroscopy to the SEI. The approach is demonstrated on reduced-graphene oxide (rGO) cycled in Li-ion cells in natural abundance and $^{13}$C-enriched electrolyte solvents. Our results indicate that DNP enhances the signal of outer SEI layers, enabling detection of natural abundance $^{13}$C spectra from this component of the SEI at reasonable timeframes. Furthermore, $^{13}$C-enriched electrolytes measurements at 100K provide ample sensitivity without DNP due to the vast amount of SEI filling the rGO pores, thereby allowing differentiating the inner and outer SEI layers composition. Developing this approach further will benefit the study of many electrode materials, equipping ssNMR with the needed sensitivity to efficiently probe the SEI.

**TOC GRAPHICS**

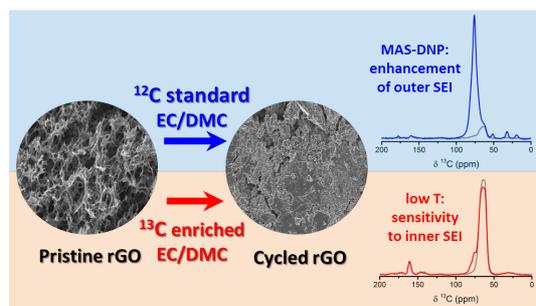

The performance and cycle life of many electrode materials used in rechargeable batteries depend on the formation of a stable and passivating surface layer called the solid electrolyte interphase (SEI)[1,2]. The formation of this layer, as a result of electrolyte reduction (on the anode) or oxidation (on the cathode), is a complex process which leads to irreversible capacity loss and can impede ion transport to the electrode, thus affecting the rate performance of the cell. The properties of the SEI and its effect on the cell's performance depend on the chemistry of the surface layer formed, namely its chemical composition and structure. Since this surface layer is often a few nanometers thick, disordered and heterogeneous (composed of amorphous mixture of organic and inorganic phases), it is challenging to characterize by conventional structural tools such as X-ray diffraction. Instead, techniques such as X-ray photoelectron spectroscopy[3] (XPS), infrared[4] (IR) and mass spectrometry[5] (MS) are commonly used to obtain a compositional map of the SEI. Another approach that is not as commonly applied in the study of the SEI is solid state nuclear magnetic resonance (ssNMR) spectroscopy, which is advantageous as it is a quantitative and chemically specific approach. Moreover, it can potentially provide structural information by employing a wide range of nuclear magnetic interactions and correlations between nuclei. However, the main drawback of ssNMR is its low sensitivity, which generally restricts its use in the study of SEI layers to the detection of nuclei with high gyromagnetic ratio and natural abundance such as $^1$H, $^{19}$F, $^{31}$P and $^7$Li[6–9], while other, lower sensitivity, nuclei such as $^{13}$C and $^{17}$O are only accessible via the use of expensive isotope enrichment and even then often require long experimental times[10–13].

Given the wealth of information that can be extracted from the $^{13}$C NMR spectra of the organic components of the SEI[10,12,13], which have a crucial role in ion transport to the electrode and providing elasticity, methods for improving the sensitivity of detection of these components are particularly important. Here we demonstrate one approach to overcome these issues by using low temperature and dynamic nuclear polarization (DNP), the latter being a technique in which the large polarization of unpaired electrons is transferred to surrounding coupled nuclear spins by microwave (MW) irradiation. DNP in combination with magic angle spinning (MAS) ssNMR is typically achieved by introducing nitroxide biradicals into the system of interest, cooling the resultant mixture to cryogenic temperatures (about 100K) and performing the MAS-ssNMR experiment while continuously irradiating the sample with MWs. This results in substantial NMR signal enhancements (up to two fold increase to date) and significant reduction in experiment time, opening the way for studies of systems that were so far limited or completely out of reach for NMR[14–16], as well as providing sufficient sensitivity for performing multidimensional experiments needed for chemical assignment and structural investigations[17,18].

The aim of this work is to explore the applicability of MAS-DNP for probing the composition of the organic components in the SEI. The approach is demonstrated on reduced graphene oxide (rGO) anodes following the first electrochemical cycle in half cell configuration in a lithium ion cell. Graphene has many desirable properties such as good electrical conductivity (about 2000S/cm), mechanical strength and extremely high surface areas (about 1500m$^2$/g)[19], therefore many have suggested that it can be used in energy storage devices[20]. In this work, the high surface area is a result of the hierarchical macroporous structure of the rGO (see scanning electron microscope, SEM, image in **Figure 1a**), which also allows the material to be cycled without the addition of binders or conductive carbon. However, its surface area also leads to large irreversible capacities (2000mAh/g on the first cycle, **Figure 1b**) associated with electrolyte degradation which leads to cell failure[21]. Thus, it is our opinion that rGO is a good model compound for SEI studies, and an ideal material with which to develop a surface sensitive NMR approach for characterizing the carbon-containing degradation products formed on the basal planes of graphitic structures; the material as currently formulated does not make a good anode material.

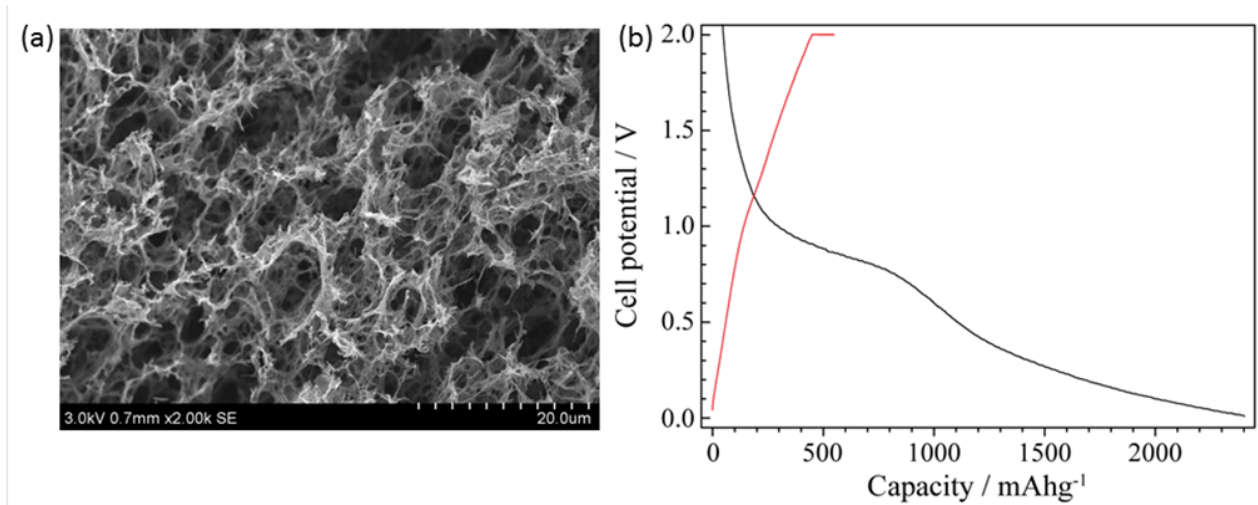

**Figure 1** (a) SEM image of pristine rGO electrode. (b) Typical voltage – capacity profile of a cell containing rGO cycled at a galvanostatically (lithiated during discharge in black and delithiated on charge in red) vs. Li metal in 1M $LiPF_6$ ethylene carbonate (EC)/ dimethyl carbonate (DMC) electrolyte with a current of 50mA/g.

A critical step in the application of MAS-DNP is sample preparation[22]. Three approaches are generally taken to introduce nitroxide biradicals into the system of interest: (i) dissolving the material of interest in a glass forming radical solution (glass forming solvent is required to prevent aggregation of the radicals) (ii) impregnating a porous material with the radical solution and (iii) impregnating a nonporous material such that the radical solution uniformly wets its surface[18]. Either aqueous or organic (non-aqueous) solvents can be used, depending on their compatibility with the material of interest, however, organic solvents were shown to have a strong effect on the enhancement (for example solvents containing methyl groups result in poor performance) with halogenated solvents leading to superior results so far[23]. Furthermore, the detection of $^{13}C$ NMR in the presence of an organic solvent adds complexity to the experiment due to the presence of the large solvent signal. The effect of the solvent on the SEI structure and stability, and the stability of the DNP radicals are important considerations that need to be addressed; for example, we cannot use water-based solvents and organic solvents need to be carefully checked with respect to SEI solubility. In the case of rGO we found that impregnating unrinsed cycled rGO electrodes with a minimal amount of 16mM TEKPol[24] solution in a mixture of partially deuterated tetrachloroethane (TCE) and deuterated chloroform (see experimental section) was a good compromise between achieving significant signal enhancement from the surface layer on cycled rGO and the dynamic range of the detection in the presence of the dominant solvent resonance. In addition, due to the conductive nature of rGO and the small sample harvested from cycled cells (5-7mg), samples were diluted with KBr powder (20-30mg) prior to the addition of about 7μl of radical solution The mixture, packed in a 3.2mm sapphire rotor, was then spun outside the NMR magnet and inserted into the MAS-DNP probe, which was kept at 100K. As the goal of this work is to examine the feasibility and limitations of the DNP experiment the samples were handled in air. Due to the reactive nature of the SEI we expect that air exposure will lead to some changes in the SEI composition and these were assessed by preliminary experiments on samples handled in a $N_2$ filled glove box.

A comparison of the signals obtained with and without MW irradiation is shown in **Figure 2**. The $^1$H NMR Hahn echo spectrum (**Figure 2a**) shows that the MW irradiation significantly enhances the $^1$H TCE solvent signal (a ratio of about 74 between the 'MW on' and 'MW off' experiments) while other contributions from the rGO are not enhanced or are masked by the dominant solvent signal and the poor spectral resolution achieved at the slow spinning frequency used (8kHz). The $^{13}$C NMR spectra (**Figure 2b(i)**) were obtained using cross-polarization (CP) from surrounding $^1$H nuclei. In the 'MW off' spectrum (green), two resonances are identified at 65 and 161ppm assigned to ethylene carbonate (EC) from residual electrolyte trapped in the rGO pores as well as possible SEI products such as alkyl carbonates and polyethylene oxide (PEO) type oligomers[13]. In addition, minor signal from the TCE solvent is observed at about 75ppm. The minor contribution of the solvent (about 50% of which is protonated) compared to the dominant EC and SEI contributions indicate it is unlikely to dissolve the SEI. When the MW irradiation is turned on, the solvent signal is substantially enhanced while only a minor change is observed in the signal intensity of the 65 and 161ppm EC resonances. However, at least four new $^{13}$C resonances can be detected at 177, 51, 32 and 19ppm which were absent from the 'MW off' spectrum. The comparison experiment (**Figure 2b(i)**) was performed with 200 transients and took about 23 minutes. An optimized $^1$H-$^{13}$C CP MAS-DNP spectrum is shown in **Figure 2b(ii)** acquired with longer CP

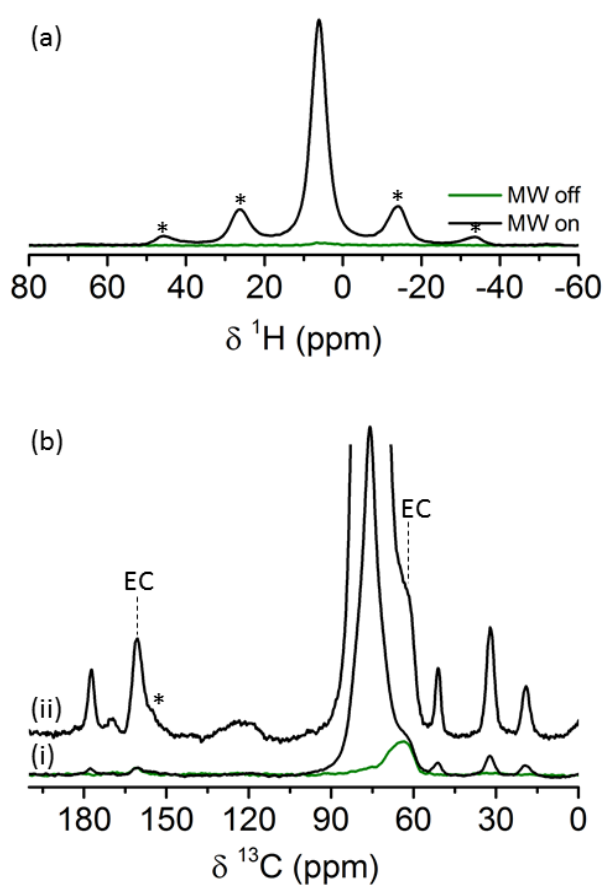

**Figure 2** (a) $^1$H rotor synchronized Hahn echo spectra of a cycled rGO sample impregnated with a radical solution (see experimental details) acquired with and without MW irradiation. (b) (i) MW on (black)/off (green) $^1$H-$^{13}$C cross polarization spectra of the same sample acquired with a contact time of 1ms and 200 transients. (ii) The enlarged spectrum was acquired with a contact time of 3ms and 3072 transients. Asterisks are used to label spinning sidebands.

contact time and 3072 transients (acquired in about 8 hours). Nine resonances can now be clearly identified, the two from EC and alkyl carbonates (65 and 161 ppm) and -OCH$_2$CH$_2$O- containing species (broad resonance at 65 ppm), three new (weaker) resonances at 183, 170 and 122ppm, and the four resonances observed previously (at 177, 51, 32 and 19ppm); these resonances are from the different functional groups in the organic phases of the SEI on rGO (discussed below), the SEI also trapping EC. A reference experiment performed at similar conditions on a pristine rGO electrode did not reveal any signals beyond the solvent resonances (see **Figure S1** in supporting information).

The signal enhancement obtained under these experimental conditions was sufficient to allow the acquisition of two-dimensional (2D) correlation experiments – $^1$H-$^{13}$C heteronuclear correlation (acquired in 20 minutes) and a $^{13}$C-$^{13}$C homonuclear correlation (using radio frequency driven recoupling, RFDR[25], acquired in 11 hours) shown in **Figure S2 and S3**. The heteronuclear correlation reveals at least three types of $^1$H environments, in addition to the TCE solvent, at 4.6ppm correlating with the two EC carbon resonances, at about 3.8ppm correlating with the carbon resonance at 51ppm and at about 3ppm correlating with the carbon resonances at 32 and 177ppm. The homonuclear correlation did not reveal any significant cross peaks (with a mixing time of 10ms), presumably because of the low natural abundance (NA) of $^{13}$C (and hence low probability of nearby $^{13}$C spins), however, it does indicate the feasibility of acquiring such spectra in a reasonable time and with sufficient S/N.

It is important to note that these experiments were performed on rGO electrodes that were cycled in NA $^{13}$C electrolyte solvents (1:1 EC/DMC) resulting in NA decomposition products in the SEI. Measurements performed at room temperature (RT) on similar samples did not reveal any signals from the SEI. When cells were cycled with $^{13}$C enriched electrolyte solvents, more than a day is required to achieve sufficient S/N in standard $^1$H-$^{13}$C CP spectra (**Figure 3a-b**) acquired at RT. Even with a day-long acquisition the main signals detected are of residual electrolyte solvents (and SEI products resonating in the same region) with some contributions from the SEI resonating at 177, 51 and 25ppm in a sample cycled in $^{13}$C$_3$-DMC enriched electrolyte (**Figure 3a**) and at 169 and 21ppm when cycled in $^{13}$C$_3$-EC enriched mixture (**Figure 3b**). Interestingly, the SEI signals detected from these samples are much broader than those acquired in the DNP experiments. Spectral broadening can be a result of fast spin-spin relaxation ($T_2$) and/or heterogeneity in the sample leading to a wide distribution of chemical shifts. Since heterogeneity in the sample would lead to broadening at both RT and low T experiments it cannot explain the broadening in RT measurements. This may indicate that relaxation due to increased dynamics of some of the SEI components is the source of broadening at RT.. An additional indication of motion at room temperature is given by the much shorter contact times used (100-500μs at 298K compared to a few ms at 100K) for achieving efficient CP transfer, the use of shorter contact times reflecting short rotating frame relaxation times of the protons ($T_{1\rho}$). In addition, significant broadening in the RT $^{13}C_3$-EC spectrum would be expected from residual $^{13}C$-$^{1}H$ coupling as this spectrum was acquired at 15kHz MAS with no $^{1}H$ decoupling due to probe arcing; that any signal is observable from of the functional groups under these conditions, again strongly suggests mobility.

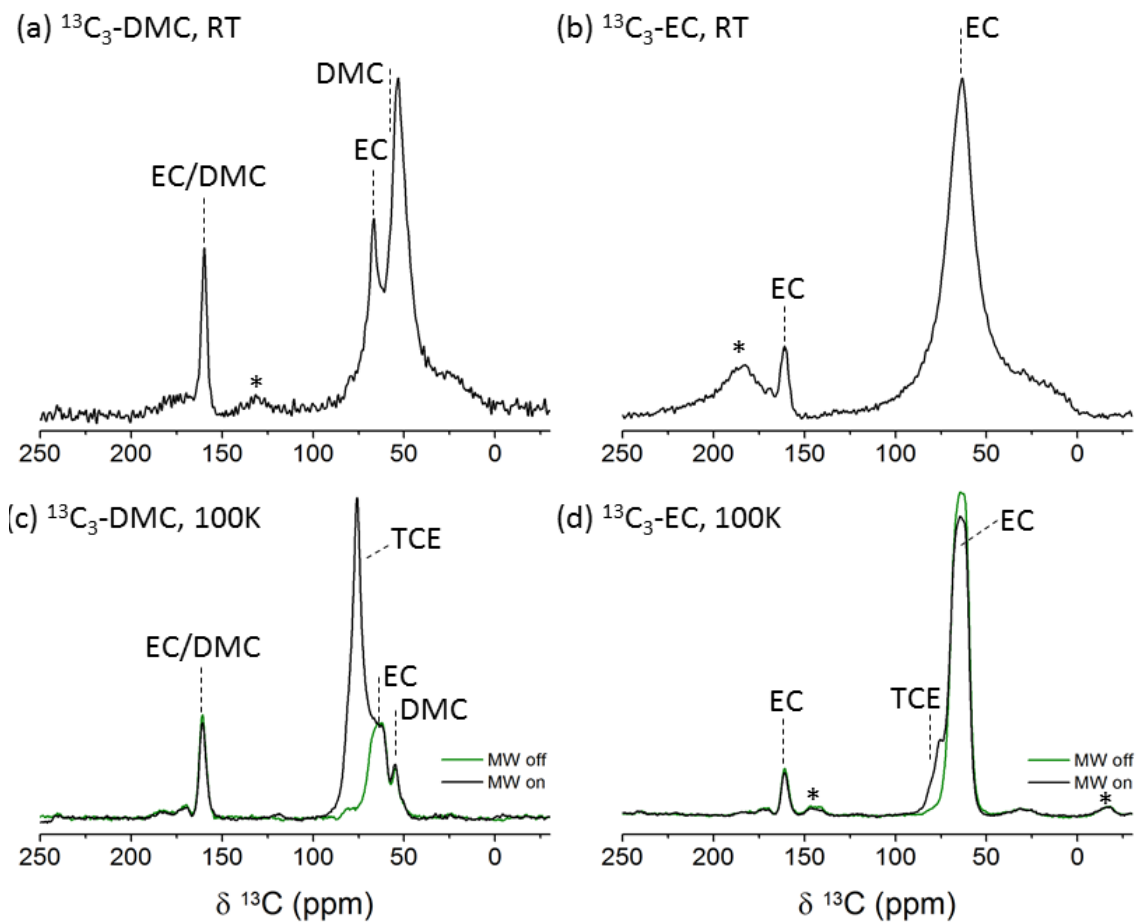

**Figure 3** $^1$H-$^{13}$C CP spectra of rGO electrodes cycled in $^{13}$C enriched electrolyte solvents containing either $^{13}$C$_3$-DMC (and NA EC) (a,c) or EC (and NA DMC) (b,d) and acquired at RT on a 11.4T magnet (a,b) or 100K on 9.4T DNP – NMR with and without MW (c,d). The contact times and number of transients used were (a) 0.1ms, 8272 (b) 0.5ms, 7504 (c) 1ms, 160 and (d) 0.5ms, 128. Relaxation delays varied between 7-10s. Spinning sidebands are marked with asterisks.

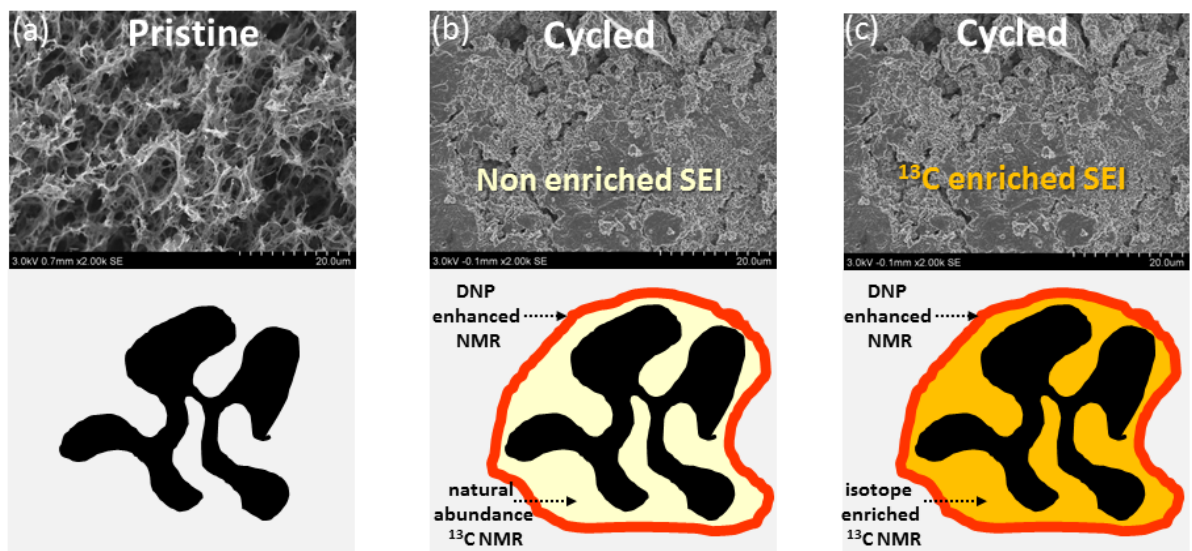

**Figure 4** SEM images of the rGO electrode before (a) and after cycling (b,c) with cartoons (the rGO structure is represented by the black shape) illustrating regions of the SEI which are detected in the various NMR experiments. The SEM image after cycling was taken from a sample cycled in na electrolyte and was duplicated for illustrative reasons.

Samples that were cycled in $^{13}$C enriched electrolytes were also characterized by MAS-DNP as they allow the decomposition products of EC and DMC to be differentiated as was recently demonstrated in silicon anodes[12,13]. CP spectra acquired at 100K from rGO cycled in either $^{13}$C$_3$-DMC or $^{13}$C$_3$-EC enriched electrolytes are shown in **Figure 3c-d**, respectively, with and without MW irradiation. Surprisingly, only the TCE solvent signal is enhanced when the MW irradiation is turned on and there is no enhancement of the electrolyte or SEI resonances (the EC signal is even slightly reduced probably due to the slight increase in temperature when the MW is turned on, see supporting info **Figure S4**). While the lack of enhancement was not expected, it can be rationalized if we consider the structure of the rGO electrode before and after cycling. The pristine rGO with its open porous framework (**Figure 4a** and cross-sectional SEM in **Figure S5a**) provides plenty of surface area for unhindered electrolyte decomposition. Upon lithiation, and without limiting the discharge voltage/capacity, electrolyte decomposition products accumulate within the pores until these become severely clogged (**Figure 4b** and cross-sectional SEM in **Figure S5b**). When the nitroxide radicals are added to the samples they can only access the external surface area of the covered rGO electrodes. Thus DNP enhancements can only be expected from the outer layer of the degradation products (and to a lesser extent from fresh surfaces that may be created by grinding the sample during the sample preparation), which is only a small fraction of the total amount of SEI formed, a much larger fraction being buried within the rGO pores. For samples cycled in non-enriched electrolyte solvents, which contain the 1% NA NMR active $^{13}$C isotope, this results in a weak NMR signal from the entire SEI. When the MW irradiation is on, a significant signal enhancement of the non- enriched SEI products in the outer layers is observed (highlighted in **Figure 4b**). However, when the electrolyte solvents and therefore SEI are enriched with $^{13}$C isotope, the majority of the SEI in the clogged pores becomes detectable by NMR at 100K, resulting in only a negligible overall enhancement of the signal due to the DNP enhancement of the outer SEI layers (**Figure 4c**). We note that our ability to enhance the signal from SEI components deposited directly on the electrode's surface may also be limited by shortening of longitudinal relaxation times caused by the conductive graphene (which will not affect the outer layers). Such effects may also explain the lack of signal enhancement from the pristine rGO electrode (**Figure S1**). Future experiments on samples that are only partially clogged by the SEI will be performed to examine these effects. Furthermore, the results suggest that very little, if any SEI dissolves into the TCE solvent, under the conditions used here as a signal enhancement from any dissolved component would be expected.

Table 1 A list of the $^{13}$C resonances detected in the CP experiments on various samples assigned to carbon functional groups and possible SEI components.

| Functional group | Possible Assignment | NA DNP (surface of SEI) | NA DNP no air exposure | $^{13}$C DMC 100 K (bulk) | $^{13}$C EC 100 K (bulk) | $^{13}$C DMC RT | $^{13}$C EC RT |
|---|---|---|---|---|---|---|---|
| R$^1$**C**O$_2$Li/H | Li formate/ oxalate | 183 | | 183 | 183 | | |
| R$^2$**C**O$_2$Li/H | Li acetate | 177 | | 177 | 177 | 173 | |
| R**C**O$_3$ | Li$_2$CO$_3$ | 170 | Weak | 170 | 170 | | 169 |
| RO**C**O$_2$R' | EC/DMC Lithium ethyl/ methyl carbonate (LEC/LMC) Lithium ethylene/butylene di-carbonate (LEDC/LBDC) | 161 | 156 | 161 | 161 | 160 | 161 |
| R**C**=**C**R' C(OR)$_4$, **C**R$_2$(OR')$_2$ | polyvinyl carbonate | 122 | 126 | 121 | | | |
| (RO**C**H$_2$)$_2$ | EC (CH$_2$) PEO-type oligomers LBDC/LEDC | 63 | 65 | 62-67 | 62-67 | 67-63 | 64 |
| **C**H$_3$OCO$_2$R | DMC (CH$_3$) | | | 55 | | 53-48 | |
| **C**H$_3$-OCO$_2$R or **C**H$_3$-OR | LMC MeOLi | 51 | 51 | 51 | | 53-48 | 53 |
| | | | 40 | | | | |
| R**C**H$_2$**C**H$_2$R' | LBDC Li Succinate | 32 | 25,31 | 25,32 (weak) | 30 | 25 (weak) | 30 |
| **C**H$_3$**C**H$_2$R (R=heteroatom or unsaturated carbon). | Li Acetate LEC | 19 | | 19 | | | |

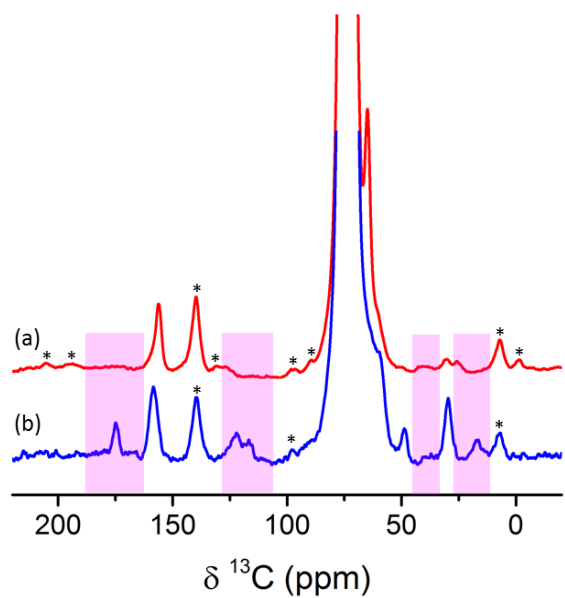

**Figure 5** $^1$H-$^{13}$C CP spectra of an rGO electrode cycled with natural abundance electrolyte and prepared for DNP experiments (a) under inert $N_2$ atmosphere, and (b) in air. DNP experiments were performed on a 14.1T spectrometer, with a contact time of 2ms, relaxation delay of 10 s and 800 (a) and 256 (b) transients. The main differences between the two samples are highlighted. Spinning sidebands are marked with asterisks.

In order to assess the effect of air exposure on the signals detected from the SEI preliminary experiments were performed on samples packed for DNP experiments in a nitrogen filled glove box. A comparison of the $^{13}$C spectra acquired from samples that were cycled in NA electrolyte with and without air exposure is shown in **Figure 5** (full spectrum shown in **Figure S6** and detected resonances listed in **Table 1**) and reveals several differences: The inert sample does not contain any significant contribution from carboxylate groups (resonating above 175 ppm) or $Li_2CO_3$ (around 170 ppm), the $sp^2$ type carbons resonate at 126ppm (and shift to 122ppm with air exposure) and the aliphatic region contains resonances at 51, 40, 31 and 25 ppm, compared to 51, 30 and 17 ppm following air exposure. We note that the products observed with air exposure were consistently detected across several samples (**Figures 5, 4** and **S7**), one of which was acquired using a different polarizing radical solution (see experimental details in the SI). Across all samples we observe variations in the amount of residual EC which is caused by differences in the amount of electrolyte initially added to the cell, contributions from EC in pieces from the glass fiber separator caught on the rGO particles and the extent of sample drying prior to the NMR measurements. Another control experiment was performed on an electrode that was measured before the addition of the radical solution (**Figure S8**) displaying at least five of the resonances that are more clearly observed in the DNP enhanced measurement. These indicate that the detected changes in the SEI composition are due to reactivity with oxygen, moisture and carbon dioxide and that the DNP approach equips us with the needed sensitivity to detect them.

Keeping in mind the effects of air exposure, we turn to analyze the differences in composition observed between the NA, $^{13}$C-DMC and $^{13}$C-EC spectra which allow us to distinguish the products formed from DMC and EC reduction (**Figure 6**). The different resonances and their possible assignment are listed in Table 1 supported by previous extensive studies on the Si system[12,13]. Since the resonances in the 20 ppm area, 51 and 122 ppm are only observed in samples cycled in $^{13}$C enriched DMC electrolyte (as was also observed in the RT NMR experiments) they must originate from DMC decomposition. These resonances can arise due to decomposition products such as lithium acetate and lithium methoxide (at 20 and 51ppm, respectively),with acetate possibly forming due to reactivity of alkoxy species with air (indicated by the pronounced appearance of the resonances around 20 and 177 ppm and the decrease around 25 ppm in figure 5b compare to 5a). Resonances in the sp$^2$ region can be assigned to unsaturated polycarbonate; mixtures of alkoxy carbons such as those found in acetals, orthoesters and orthocarbonates have also been proposed to be responsible for resonances with similar chemical shifts[10] to that of the broad signal at 122ppm.

As the enhancement achieved by DNP seems to be limited to the external SEI layer covering the electrode it can provide insight into the structure of the SEI. This can be used to distinguish the composition in the inner pores close to the carbon surface, dominating the spectra of $^{13}$C enriched samples, from the outer layers enhanced by DNP. Here the DNP enhanced spectrum has a much higher contribution from aliphatic carbons (19-32ppm) as well as higher carboxylate content resonating at 177ppm.. The higher intensity of the resonance at 183ppm in the $^{13}$C enriched samples implies the bulk contains more carboxylate species such as formate or oxalate, although we note that further NMR experiments, done under inert conditions, are required before definitive assignments can be made. Furthermore, the signal intensity in the DNP experiments should be interpreted with care as it depends on the transfer of polarization from the frozen solvent molecules to the sample via spin-diffusion[26] which is not necessarily uniform. Significant differences between the various samples are also observed in the region of 62-

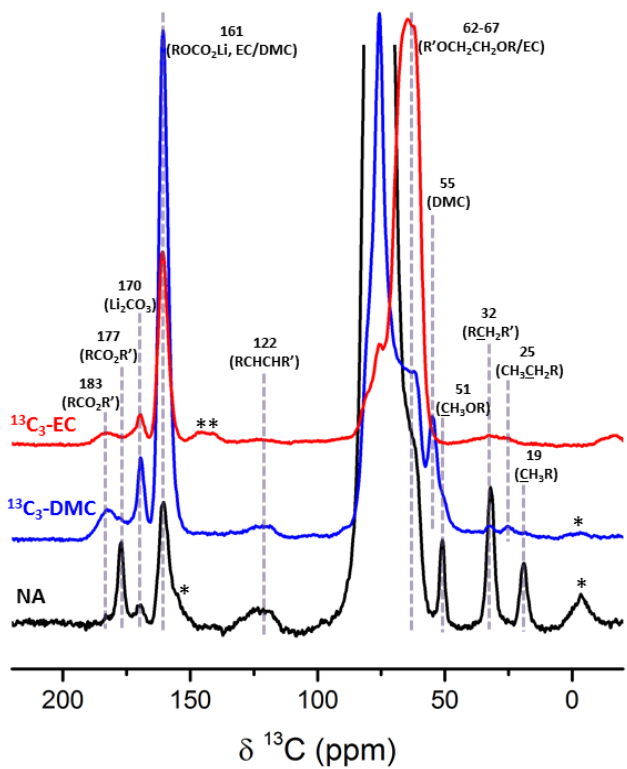

**Figure 6** $^1H$-$^{13}C$ CP spectra of rGO electrode cycled in NA electrolyte (black, 3072 transients) and in electrolyte with $^{13}C_3$-DMC (blue, 3072 transients) and $^{13}C_3$-EC (red, 800 transients). All spectra were acquired with a contact time of 3ms, relaxation delay of 7s and MW irradiation.

67ppm- which contains contributions from EC, alkyl carbonates and PEO species. The low resolution in the 1D experiment, especially in this crowded region, does not allow distinguishing the various components. However, the increased sensitivity offered by the low temperature and DNP enhancement can potentially be used for assigning these resonances via 2D correlation experiments, as was recently demonstrated for the SEI on Si[13].

S6)

Our results demonstrate the increased sensitivity offered by the MAS-DNP approach to the surface layer of the SEI. Significant signal enhancement can be achieved in reasonable experiment time with no isotope enrichment, suggesting that further assignments of the various SEI components and structural studies are possible via multidimensional experiments. In the case of rGO, due to its extremely large porosity, the use of $^{13}C$ labeled electrolytes in combination with the low temperature experiments result in substantial signal from the internal layers which form the majority of the SEI. In this case, the DNP enhancement of the outermost surface layers becomes negligible but can be used to compare the inner and outer SEI layers. Our experiments suggest that the organic components of the SEI cannot be viewed as rigid salt crystals, but rather the various functional groups display considerable mobility/dynamics (at RT). The fluxionality of some of the SEI components will likely play an important role in controlling $Li^+$ transport through the SEI. These measurements also provide clear evidence for the formation of SEI products containing saturated carbon groups (as seen in the SEI of Si), which must originate from extensive radical reactions involving more than one electrolyte molecule. We expect further development of this approach will benefit the study of the surface layers formed in many electrode materials, equipping solid state NMR with the needed sensitivity to efficiently probe the chemical and structural compositional changes in the SEI layer.

**Experimental Section**

Graphene oxide (GO) was synthesized by a modified Hummer's method, the exact details of the synthesis are given elsewhere[27]. The resulting GO solution was hydrothermally reduced at 180°C for 6 hours and then freeze-dried for 48 hours to provide a porous rGO. The rGO was further characterized by solid state $^{13}$C CPMG NMR[28] and Raman spectroscopy (see Figure S9). The prepared rGO was further dried under vacuum at 150 °C and then immediately transferred into an Ar glove box and assembled into coin cells. Further details of electrochemical processes are provided in the SI.

NMR experiments were performed on cycled rGO powder extracted from the coin cell and dried under vacuum for at least 15 hours in the glove box pre-chamber. The resulting unrinsed and dry powder was packed in the glove box for MAS NMR measurements or thoroughly mixed with dry KBr powder and packed in air for MAS-DNP. For the latter about 7μl of 16mM TEKPol in TCE (20% $CDCl_3$, 24% $d_4$-TCE, 56% TCE) were centrifuged into the rotor (see SI).

**Acknowledgements**

The work was supported by a research grant from Dana and Yossie Hollander and the Alon fellowship from Israel council of higher education (ML). This project has received funding from the European Unions's Horizon 2020 research and innovation programme under grant agreement No. 696656 – GrapheneCore1 (GK and CPG). We thank Dr. Wanjing Yu (Central South University, China) for graphene synthesis and related discussions. GK thanks Dr. Duhee Yoon (Cambridge Graphene Centre) for Raman measurements and helpful discussions. The research is made possible in part by the historic generosity of the Harold Perlman family. We thank Dr. Frederic

Mentink-Vigier for helpful suggestions. DNP experiments at 14.1 T were performed at the DNP MAS NMR Facility at the University of Nottingham, with thanks to the EPSRC for funding of pilot studies (EP/L022524/1).

Supporting Information Available: Experimental details on electrochemistry, sample preparation for NMR and DNP experiments. Additional results from pristine and cycled rGO samples, 2D $^1$H-$^{13}$C correlation and 2D $^{13}$C-$^{13}$C homonuclear correlation, temperature calibration, cross sectional SEM images, results from samples that were not exposed to air and characterization of pristine rGO by NMR and RAMAN spectroscopy..

# References

(1) Peled, E. The Electrochemical Behavior of Alkali and Alkaline Earth Metals in Nonaqueous Battery Systems—The Solid Electrolyte Interphase Model. *J. Electrochem. Soc.* **1979**, *126* (12), 2047.

(2) Xu, K. Electrolytes and Interphases in Li-Ion Batteries and Beyond. *Chem. Rev.* **2014**, *114*, 11503–11618.

(3) Philippe, B.; Dedryvère, R.; Allouche, J.; Lindgren, F.; Gorgoi, M.; Rensmo, H.; Gonbeau, D.; Edström, K. Nanosilicon Electrodes for Lithium-Ion Batteries: Interfacial Mechanisms Studied by Hard and Soft X-Ray Photoelectron Spectroscopy. *Chem. Mater.* **2012**, *24* (6), 1107–1115.

(4) Nie, M.; Abraham, D. P.; Chen, Y.; Bose, A.; Lucht, B. L. Silicon Solid Electrolyte Interphase (SEI) of Lithium Ion Battery Characterized by Microscopy and Spectroscopy. *J. Phys. Chem. C* **2013**, *117* (26), 13403–13412.

(5) Lu, P.; Li, C.; Schneider, E. W.; Harris, S. J. Chemistry, Impedance, and Morphology Evolution in Solid Electrolyte Interphase Films during Formation in Lithium Ion Batteries. *J. Phys. Chem. C* **2014**, *118* (2), 896–903.

(6) Meyer, B. M.; Leifer, N.; Sakamoto, S.; Greenbaum, S. G.; Grey, C. P. High Field Multinuclear NMR Investigation of the SEI Layer in Lithium Rechargeable Batteries. *Electrochem. Solid-State Lett.* **2005**, *8* (3), A145.

(7) Dupré, N.; Cuisinier, M.; Guyomard, D. Electrode/Electrolyte Interface Studies in Lithium Batteries Using NMR. *Interface* **2011**, No. Fall, 61–67.

(8) Cuisinier, M.; Dupré, N.; Martin, J.-F.; Kanno, R.; Guyomard, D. Evolution of the LiFePO4 Positive Electrode Interface along Cycling Monitored by MAS NMR. *J. Power Sources* **2013**, *224*, 50–58.

(9) Murakami, M.; Yamashige, H.; Arai, H.; Uchimoto, Y.; Ogumi, Z. Direct Evidence of LiF Formation at Electrode/Electrolyte Interface by 7Li and 19F Double-Resonance Solid-State NMR Spectroscopy. *Electrochem. Solid-State Lett.* **2011**, *14* (9), A134.

(10) Leifer, N.; Smart, M. C.; Prakash, G. K. S.; Gonzalez, L.; Sanchez, L.; Smith, K. a.; Bhalla, P.; Grey, C. P.; Greenbaum, S. G. 13C Solid State NMR Suggests Unusual Breakdown Products in SEI Formation on Lithium Ion Electrodes. *J. Electrochem. Soc.* **2011**, *158* (5), A471.

(11) Delpuech, N.; Dupré, N.; Mazouzi, D.; Gaubicher, J.; Moreau, P.; Bridel, J. S.; Guyomard, D.; Lestriez, B. Correlation between Irreversible Capacity and Electrolyte Solvents Degradation Probed by NMR in Si-Based Negative Electrode of Li-Ion Cell. *Electrochem. commun.* **2013**, *33*, 72–75.

(12) Michan, A. L.; Leskes, M.; Grey, C. P. Voltage Dependent Solid Electrolyte Interphase Formation in Silicon Electrodes: Monitoring the Formation of Organic Decomposition Products. *Chem. Mater.* **2016**, *28* (1), 385–398.

(13) Michan, A. L.; Divitini, G.; Pell, A. J.; Leskes, M.; Ducati, C.; Grey, C. P. Solid Electrolyte Interphase Growth and Capacity Loss in Silicon Electrodes. *J. Am. Chem. Soc.* **2016**, *138* (25), 7918–7931.

(14) Sangodkar, R. P.; Smith, B. J.; Gajan, D.; Rossini, A. J.; Roberts, L. R.; Funkhouser, G. P.; Lesage, A.; Emsley, L.; Chmelka, B. F. Influences of Dilute Organic Adsorbates on the Hydration of Low-Surface-Area Silicates. *J. Am. Chem. Soc.* **2015**, *137* (25), 8096–8112.

(15) Lilly Thankamony, A. S.; Lion, C.; Pourpoint, F.; Singh, B.; Perez Linde, A. J.; Carnevale, D.; Bodenhausen, G.; Vezin, H.; Lafon, O.; Polshettiwar, V. Insights into the Catalytic Activity of Nitridated

Fibrous Silica (KCC-1) Nanocatalysts from 15 N and 29 Si NMR Spectroscopy Enhanced by Dynamic Nuclear Polarization. *Angew. Chemie Int. Ed.* **2015**, *54* (7), 2190–2193.

(16) Johnson, R. L.; Perras, F. A.; Kobayashi, T.; Schwartz, T. J.; Dumesic, J. A.; Shanks, B. H.; Pruski, M. Identifying Low-Coverage Surface Species on Supported Noble Metal Nanoparticle Catalysts by DNP-NMR. *Chem. Commun.* **2016**, *52* (9), 1859–1862.

(17) Ni, Q. Z.; Daviso, E.; Can, T. V; Markhasin, E.; Jawla, S. K.; Swager, T. M.; Temkin, R. J.; Herzfeld, J.; Griffin, R. G. High Frequency Dynamic Nuclear Polarization. *Acc. Chem. Res.* **2013**, *46* (9), 1933–1941.

(18) Rossini, A. J.; Zagdoun, A.; Lelli, M.; Lesage, A.; Copéret, C.; Emsley, L. Dynamic Nuclear Polarization Surface Enhanced NMR Spectroscopy. *Acc. Chem. Res.* **2013**, *46* (9), 1942–1951.

(19) Raccichini, R.; Varzi, A.; Passerini, S.; Scrosati, B. The Role of Graphene for Electrochemical Energy Storage. *Nat. Mater.* **2014**, *14* (3), 271–279.

(20) Xu, C.; Xu, B.; Gu, Y.; Xiong, Z.; Sun, J.; Zhao, X. S. Graphene-Based Electrodes for Electrochemical Energy Storage. *Energy Environ. Sci.* **2013**, *6* (5), 1388.

(21) Vargas, Ó.; Caballero, Á.; Morales, J.; Rodríguez-Castellón, E. Contribution to the Understanding of Capacity Fading in Graphene Nanosheets Acting as an Anode in Full Li-Ion Batteries. *ACS Appl. Mater. Interfaces* **2014**, *6* (5), 3290–3298.

(22) Lee, D.; Hediger, S.; De Paëpe, G. Is Solid-State NMR Enhanced by Dynamic Nuclear Polarization? *Solid State Nucl. Magn. Reson.* **2015**, *66–67*, 6–20.

(23) Zagdoun, A.; Rossini, A. J.; Gajan, D.; Bourdolle, A.; Ouari, O.; Rosay, M.; Maas, W. E.; Tordo, P.; Lelli, M.; Emsley, L.; et al. Non-Aqueous Solvents for DNP Surface Enhanced NMR Spectroscopy. *Chem. Commun.* **2012**, *48* (5), 654.

(24) Zagdoun, A.; Casano, G.; Ouari, O.; Schwarzwälder, M.; Rossini, A. J.; Aussenac, F.; Yulikov, M.; Jeschke, G.; Copéret, C.; Lesage, A.; et al. Large Molecular Weight Nitroxide Biradicals Providing Efficient Dynamic Nuclear Polarization at Temperatures up to 200 K. *J. Am. Chem. Soc.* **2013**, *135* (34), 12790–12797.

(25) Bennett, A. E.; Rienstra, C. M.; Griffiths, J. M.; Zhen, W.; Lansbury, P. T.; Griffin, R. G. Homonuclear Radio Frequency-Driven Recoupling in Rotating Solids. *J. Chem. Phys.* **1998**, *108* (22), 9463.

(26) Van Der Wel, P. C. a; Hu, K. N.; Lewandowski, J.; Griffin, R. G. Dynamic Nuclear Polarization of Amyloidogenic Peptide Nanocrystals: GNNQQNY, a Core Segment of the Yeast Prion Protein Sup35p. *J. Am. Chem. Soc.* **2006**, *128*, 10840–10846.

(27) Liu, T.; Leskes, M.; Yu, W.; Moore, A. J.; Zhou, L.; Bayley, P. M.; Kim, G.; Grey, C. P. Cycling Li-O2 Batteries via LiOH Formation and Decomposition. *Science (80-. ).* **2015**, *350* (6260), 530–533.

(28) Hung, I.; Rossini, A. J.; Schurko, R. W. Application of the Carr−Purcell Meiboom−Gill Pulse Sequence for the Acquisition of Solid-State NMR Spectra of Spin- 1 / 2 Nuclei. *J. Phys. Chem. A* **2004**, *108* (34), 7112–7120.